\documentclass{article}

\usepackage[final]{neurips_2019}

  \usepackage{natbib}
\setcitestyle{square,numbers}

\usepackage[utf8]{inputenc} 
\usepackage[T1]{fontenc}    
\usepackage{hyperref}       
\usepackage{url}            
\usepackage{booktabs}       
\usepackage{amsfonts}       
\usepackage{nicefrac}       
\usepackage{microtype}      

\title{On the Legal Compatibility of Fairness Definitions}

\author{%
  Alice Xiang \\
  Partnership on AI\\
  San Francisco, CA 94109 \\
  \texttt{alice@partnershiponai.org} \\
   \And
   Inioluwa Deborah Raji \\
   Partnership on AI \\
   San Francisco, CA 94109 \\
   \texttt{deb@partnershiponai.org} \\
   }

\begin{document}

\maketitle

\begin{abstract}
Past literature has been effective in demonstrating ideological gaps in machine learning (ML) fairness definitions when considering their use in complex socio-technical systems. However, we go further to demonstrate that these definitions often misunderstand the legal concepts from which they purport to be inspired, and consequently inappropriately co-opt legal language. In this paper, we demonstrate examples of this misalignment and discuss the differences in ML terminology and their legal counterparts, as well as what both the legal and ML fairness communities can learn from these tensions.  We focus this paper on U.S. anti-discrimination law since the ML fairness research community regularly references terms from this body of law. 
\end{abstract}

\section{Why Legal Compatibility is Important}



ML fairness literature is often presented as a step toward reducing algorithmic bias in high-stakes human systems---from the COMPAS risk assessment tool as part of the broader judicial system \cite{jung2019eliciting} 
to hypothetical loan mis-allocation scenarios affecting human livelihoods \cite{hardt2016equality}. Others have identified a gap between ML measurements and the complex reality of the socio-technical human environments in which they are deployed  \cite{selbst2019fairness,friedler2016possibility}. One key aspect of this gap is the divergence between legal and ML fairness concepts. Although the ML fairness community often uses legal terminology to motivate their research, these terms are used without a deeper understanding of their meanings.

To build frameworks of accountability and governance, ML fairness definitions must align with their corresponding legal definitions. This alignment must happen from both sides, as key lessons from the ML fairness literature should also inform how the law evolves to accommodate algorithmic bias. There are two primary legal scenarios to consider. First, to demonstrate evidence of algorithmic unfairness, ML definitions must accurately map onto their legal counterparts to establish liability. For algorithm developers to be incentivized to adopt particular bias measures, these metrics must conform to what judges, juries, or agencies would accept as evidence of discrimination or lack thereof. Second, to deploy bias correction methods in the real world, it is important that the methods themselves are not determined to be discriminatory from a legal perspective. Thus, in order for the research of the ML fairness community to have an impact on deployed systems, we must assess the legal compatibility of this work.


In this short paper, we present various examples of the current tension that exists between ML fairness and legal definitions. We go further to provide cases of legal precedent to illustrate the issues that arise from the lack of compatibility of these definitions. We focus on lessons from U.S. anti-discrimination law since the ML fairness research community regularly references terms from this body of law. In addition, we explore lessons that the legal community can learn from the ML fairness community. We hope this work encourages those from the legal and ML fairness communities to collaborate on research and initiatives that have real-world impact on the fairness of ML tools making consequential decisions within complex human systems. 



\section{Lessons from Anti-Discrimination Law for ML Fairness}

In this section, we discuss terms from anti-discrimination law that are often co-opted or misinterpreted in the ML fairness literature.

\subsection{Discrimination}
In ML literature, ``discrimination'' is often presented as an unjust correlation between protected class variables and some metric of interest, such as outcomes, false positive rates, or a similarity metric \cite{hardt2016equality, jung2019eliciting}. Perceived as an illogical phenomenon in the system, many projects aim to ``optimally adjust any learned predictor so as to remove discrimination'' \cite{hardt2016equality}. However, this view is over-simplified when considering how discrimination is defined legally. The Civil Rights Act of 1964, Fair Housing Act, Age Discrimination in Employment Act, and Americans with Disabilities Act, among other statutes, provide anti-discrimination protections in housing, employment, and other domains. These acts primarily define discrimination though motive, evidenced intent of exclusion, and causality, rather than simply outcomes. Moreover, they are highly domain-specific. Gauging this discriminatory intent is complicated, often involves specific contextual compliance checklists and references \cite{jacobs2000fair}, and cannot be easily ``removed.'' 
\subsection{Protected Class/Sensitive Attribute}
In ML, protected attributes are presented as recorded or visible traits that should not factor into a decision, such as age, race, or gender. Legally, less commonly measured attributes can also be considered, such as sexual orientation, pregnancy, and disability status. Moreover, although the ML community often refers to members of protected classes as those in ``minority and marginalized groups'' \cite{hutchinson2019interpreting}, current anti-discrimination laws operate symmetrically on the attributes of race and gender \cite{areheart2017symmetry}. In fact, many landmark anti-discrimination law cases feature white male plaintiffs arguing that anti-discrimination law protects them against policies that seek to benefit minorities \cite{primus2015visible, bakke, ricci}. As the ML community develops technical definitions for fairness, it is important for researchers to contemplate whether they would want their definitions to be used by those in the majority against those in the minority. 

\subsection{Anti-classification and Anti-subordination}

There are two major motivations for anti-discrimination law---anti-classification and anti-subordination \cite{balkin2003american}. Anti-classification is the idea that classifications based on protected class attributes are impermissible. The ML fairness community is actually quite familiar with this concept---``fairness through unawareness'' \cite{dwork2012fairness} is an instantiation of this principle, and there has been work explicitly presenting anti-classification as a potential fairness objective \cite{corbett2018measure}. 
However, anti-subordination is rarely called out as a motivation in ML fairness literature. Anti-subordination is the idea that anti-discrimination laws should serve to actively dismantle societal hierarchies between different groups, empowering historically disenfranchised and vulnerable groups, even if doing so requires unequal treatment of a particular group. Although debiasing techniques developed by the ML community might have the effect of anti-subordination \cite{lipton2018does}, this balancing of the scales of historical injustice and hierarchies is rarely explicitly acknowledged or understood as a justified goal. 
 
\subsection{Affirmative Action}

The ML fairness community has articulated a goal of ```fair affirmative action,' which guarantees statistical parity (i.e., the demographics of the set of individuals receiving any classification are the same as the demographics of the underlying population), while treating similar individuals as similarly as possible'' \cite{dwork2012fairness} and understands affirmative action to be cases in which we ``explicitly take demographics into account'' \cite{kannan2019downstream}. 
However, the legal precedent around affirmative action demonstrates this interpretation to be inaccurate, particularly when applied to the algorithmic context.

Landmark affirmative action cases have concluded that schools seeking to increase racial diversity cannot use racial quotas \cite{bakke} or point systems \cite{grutter}. The Supreme Court, however, has permitted the use of race as one of many holistic factors while evaluating candidates as individuals \cite{grutter,fisher}. In higher education, schools have dealt with this conundrum through greater opacity, seeking to be race conscious without making explicit how race factors into admissions decisions \cite{primus2015visible}. Thus, an unsophisticated direct application of these precedents to the algorithmic context would suggest that developers who want to address bias in their algorithms should be opaque in their use of protected class variables or avoid use altogether, practices that the ML community explicitly denounces as ``naive'' \cite{hardt2016equality}. 

\subsection{Disparate Treatment and Disparate Impact}

In ML literature, disparate treatment is often explained as making use of the protected attribute in the decision-making process \cite{jagielski2018differentially}, while disparate impact is understood as when ``outcomes differ across subgroups (even unintentionally)'' \cite{lipton2018does}. In this work, disparate treatment is presented as a justification for avoiding the use of protected class variables in debiasing techniques \cite{dwork2012fairness,lipton2018does}, while disparate impact is used to justify group fairness formulations \cite{feldman2015certifying, corbett2018measure,barocas2016big}. 

However, these concepts were developed with human discriminators in mind, so simply replacing the human decision-maker with an algorithmic one is often not appropriate. For instance, the key legal question for disparate treatment is whether the alleged discriminator's actions were motivated by discriminatory intent \cite{griggs}. Intent is an inherently human characteristic---an algorithm cannot possess intent by itself, so the characterization of disparate treatment as the use of protected class variables in an algorithm is highly contestable. In fact, it is unclear if the concept of disparate treatment is relevant to evaluating the non-human components of a system at all.

Although disparate impact seems more focused on outcomes, these disproportionate outcomes are still seen by some as ``smoke'' or evidence of the fire of discriminatory intent \cite{ricciscalia, primus2015visible}. Showing disproportionate outcomes is only the first step of a disparate impact case; there is only liability if the defendant cannot explain or justify these disproportionate outcomes with nondiscriminatory rationales or the plaintiff cannot establish a less discriminatory alternative that similarly achieves legitimate business goals \cite{rutherglen1987disparate}. 

Disparate treatment and disparate impact  are sometimes in tension when organizations seek to prevent disproportionate outcomes by being race conscious. Ricci v. DeStefano \cite{ricci} is a key employment discrimination case illustrating this tension. In that case, the city of New Haven voided the results of a promotion process for firefighters after realizing that the process would have only promoted white and Hispanic firefighters and no black firefighters. The Court concluded that this constituted disparate treatment since the city did not have a ``strong basis in evidence" that it would have been subjected to disparate impact liability if it had followed through with the promotions. This case has led some scholars in the ML fairness community \cite{kroll2016accountable} to conclude that correcting bias in algorithms ex post might be prohibited by law. Others in the legal community, however, \cite{kim2016data,primus2015visible} have interpreted Ricci more narrowly, concluding that it would not limit most efforts to correct for bias since the Court in this case was heavily motivated by the fact that many firefighters spent significant time and money preparing for the test, an issue that would not apply in most algorithmic contexts. 

\section{Broader Legal Lessons for ML Fairness}

Outside the specific doctrines of anti-discrimination law, there are some general legal fairness principles that can be useful to consider for algorithmic fairness.

\subsection{Intersectionality}

Intersectionality was motivated by critical race theory to humanize subjects and highlight their individual considerations in legal decision-making. For example, legal scholars discuss how black women often struggled to succeed in anti-discrimination cases if they could not establish that their experiences were reflective of discrimination against black individuals generally or female individuals generally, rather than the black women specifically \cite{crenshaw1989demarginalizing}. Similarly, ML fairness literature tends to categorize and stereotype individuals in a way that masks more specific individual-level harms, at best only considering additive combinations of unitary classes (i.e., you are white and a man) \cite{buolamwini2018gender, hoffmann2019fairness}. Even attempts at more granular algorithmic ``individual fairness'' relies on a similarity metric and minimizing the influence of protected variables \cite{dwork2012fairness}, rather than acknowledging and seeking to account for the combination of oppressive challenges faced by the individual.

\subsection{Procedural Justice}
In ML, the term ``procedural fairness'' has been co-opted to refer to identifying the input features that lead to a particular model outcome, as a proxy for the ``process'' through which the model makes its prediction \cite{grgic2018beyond}. This is a narrow and misguided view of what this legal term means. Rather than attempting to specify definitions of fair outcomes, procedural justice seeks to arrive at just outcomes through iterative processes and the close examination of the set of governance structures in place to guide individual human decision-making \cite{citron2014scored}. The goal should thus be to analyze the system surrounding the algorithm and its use, including posing key questions on contestability and due diligence, rather than scoping down into the specifics of the algorithm itself. 

\section{Lessons from ML Fairness for Law}

Finally, we address a few examples of how the legal and policy communities could learn from the ML fairness research community. 

\subsection{Measurement of Bias}

What makes algorithms unique is that in theory it is easier to empirically measure discrimination in these defined systems than their very messy human counterparts. That said, moving away from ``fairness through unawareness'' and making a case for the use of protected variables to effectively measure discrimination, although an intuitive and proven conclusion from a mathematical perspective, is still much more difficult to make a case for in a legal sense. Making the technical case for why it is important to relax restrictions on the collection and use of demographic data, and have this data accessible for external audits would inform the legal discourse on this topic.

\subsection{Narrow Domain of Application}

 The Constitution primarily protects against discrimination by government actors, so if Congress or state legislatures do not pass anti-discrimination laws against private actors in a specific domain, individuals have little legal recourse. Although ML research often discusses the fairness of ad targeting and recommendation algorithms, bias in these algorithms---although potentially immoral---is not generally illegal. For example, the ProPublica article on Facebook's advertising targeting platform, which allowed advertisers to exclude users based on their ``ethic affinity'' \cite{PPFB_1}, resulted in private class-action lawsuits and charges from the Department of Housing and Urban Development (HUD). However, these legal actions only applied to advertisements concerning housing, employment, and credit/lending, leading Facebook to make changes restricting demographic targeting for ads in these domains, but not in others \cite{PPFB_2}. ML fairness researchers often focus more broadly, reaching beyond the domains of current anti-discrimination law. As the use of ML algorithms proliferates, this in turn can open up the conversation for what new anti-discrimination protections to consider. 

\subsection{Causality}

Most anti-discrimination laws are motivated by causality, i.e. decisions cannot be made ``because of'' an individual's protected class. However, judges and policymakers are often not aware of what would indicate causality in the context of algorithms. 
For example, the HUD recently proposed a rule that would create a safe harbor from disparate impact claims for housing algorithms that do not use protected class variables (or close proxies) \cite{HUD_FB}. The rule seeks to implement the Supreme Court's decision in Texas Department of Housing and Community Affairs v. Inclusive Communities Project, Inc. \cite{texas}, which requires housing discrimination plaintiffs to ``show a causal connection between the Department's policy and a disparate impact.'' That said, from a statistical perspective, the presence or absence of protected class variables (or close proxies) in the algorithm does not indicate the presence or absence of a causal connection \cite{kusner2017counterfactual}. It is thus important for the ML research community to assist policymakers with translating causality from a legal perspective into the technical context.

\section{Conclusion}

If the goal is for ML models to operate effectively within human systems, they must be compatible with human laws.
In order for ML researchers to produce impactful work and for the law to accurately reflect the technical realities of algorithmic bias, these disparate communities must recognize each other as partners to collaborate with closely and allies to aid in building a shared understanding of algorithmic harms and the appropriate interventions, ensuring that they are compatible with real-world legal systems. 


	\bibliographystyle{plain}
	\bibliography{neurips_2019}

\end{document}